# Electrical Transport Properties of Single-Layer WS$_2$

Dmitry Ovchinnikov[1], Adrien Allain[1], Ying-Sheng Huang[2], Dumitru Dumcenco[1], Andras Kis[1*]

[1]Electrical Engineering Institute, Ecole Polytechnique Federale de Lausanne (EPFL), CH-1015 Lausanne, Switzerland

[2]Department of Electronic Engineering, National Taiwan University of Science and Technology, Taipei 106, Taiwan

*Correspondence should be addressed to: Andras Kis, andras.kis@epfl.ch

**ABSTRACT**

We report on the fabrication of field-effect transistors based on single and bilayers of the semiconductor WS$_2$ and the investigation of their electronic transport properties. We find that the doping level strongly depends on the device environment and that long *in-situ* annealing drastically improves the contact transparency allowing four-terminal measurements to be performed and the pristine properties of the material to be recovered. Our devices show n-type behavior with high room-temperature on/off current ratio of ~10$^6$. They show clear metallic behavior at high charge carrier densities and mobilities as high as ~140 cm$^2$/Vs at low temperatures (above 300 cm$^2$/Vs in the case of bi-layers). In the insulating regime, the devices exhibit variable-range hopping, with a localization length of about 2 nm that starts to increase as the Fermi level enters the conduction band. The promising electronic properties of WS$_2$, comparable to those of single-layer MoS$_2$ and WSe$_2$, together with its strong spin-orbit coupling, make it interesting for future applications in electronic, optical and valleytronic devices.

Transistor scaling issues related to increased heat dissipation due to short-channel effects clearly show the strong need for new materials with enhanced characteristics for electronics. Two-dimensional (2D) materials and transition metal dichalcogenides (TMDs) among them, are widely investigated at the moment, due to the rich diversity of mechanical, electronic and optoelectronic properties. Initiated in 2004 by the isolation of graphene,[1] which further became one of the most intensively investigated materials, the field expanded to other 2D materials such as TMDs. Despite its outstanding carrier mobility,[2] graphene lacks a bandgap, which limits its applications in electronics. TMDs on the other hand are a family of materials with a wide variety of electronic properties.[3] Typically described by the formula MX$_2$, where M stands for a transition metal (Mo, W, Nb, Ta) and X for a chalcogenide atom (S, Se, Te), this material family is composed of semiconductors (MoS$_2$, MoSe$_2$, WS$_2$, WSe$_2$, …), metals (NbTe$_2$, TaTe$_2$, …) and superconductors (NbS$_2$, NbSe$_2$, …). Among the semiconductors in the family of TMDCs, MoS$_2$ isolated in the single-layer form was first to attract attention, demonstrating high on/off current ratio and low off-state current.[4] Integrated circuits based on single-layer MoS$_2$,[4] bilayer MoS$_2$,[5] signal amplifiers[6] and non-volatile memory cells[7] were also demonstrated as well as highly flexible FET devices based on ultrathin MoS$_2$.[8] The presence of a direct band gap[9-12] in MoS$_2$ due to the 2D confinement[12] makes it interesting for applications in optoelectronics, for example in ultrasensitive[13] photodetectors.[14] At low temperatures, metal-insulator transition in single-layer MoS$_2$ devices with top-[15] and back-gates were observed.[16] Demonstration of high-performance devices in the back-gated geometry,[16,17] with electron mobilities > 60 cm$^2$V$^{-1}$s$^{-1}$ at room temperature highlighted another important property of these materials: the strong influence of the adsorbates from the atmosphere on their electronic properties.

WS$_2$ is another representative of the family of semiconducting TMDs. According to optical measurements, the single-layer form exhibits direct band gap of at least 2.0 eV.[18] Theoretical models predict that among the semiconducting 2D TMDs,[19] WS$_2$ should have the highest mobility due to the reduced effective mass. WS$_2$ could also be very interesting because of the strong spin-orbit coupling induced valence band splitting[20] which results in spin-valley coupling.[21] While the valence band splitting for MoS$_2$ is ≈150 meV, because of the larger mass of W atoms, it is almost three times larger (≈426 meV) in WS$_2$, which would allow easier observation of the valley Hall effect than in MoS$_2$.[21, 22]

The need for large-area deposition of TMDs resulted in interest in large-scale growth methods, including CVD growth of WS$_2$ in single-layer form.[23-25] All these outline the need for data on the electrical performance and transport properties of single-layer WS$_2$. Recent reports of transistors made of multilayer WS$_2$ (ref. 26) showed the potential of this material. Vertical transistors based on WS$_2$/graphene heterostructures were also integrated with flexible substrates.[27] Recent reports on liquid-gated multilayer[28] and single-layer[29] WS$_2$ gave an estimate of the mobility of charge carriers of 44 cm$^2$/Vs at room temperature.[29] Temperature-dependent transport of multilayer WS$_2$ in the insulating regime was also recently investigated on SiO$_2$ and BN substrates.[30] Temperature-dependent transport measurements on single-layer WS$_2$ with a solid gate, which are desirable for practical applications are still missing. In this letter we aim to complete the picture and provide experimental material regarding the transport properties of this material. After careful annealing in vacuum, we were able to probe both the insulating and the metallic conduction regime in this material as a function of temperature, reaching high mobilities at room temperature and observing the crossover into the metallic conduction regime at lower temperatures.

**RESULTS AND DISCUSSION**

**Device performance in air and in vacuum**. We first focus on the room temperature performance of our back-gated WS$_2$ field-effect transistors. An optical micrograph of a typical device is shown on Figure 1b. Bias voltage is applied between drain and source electrodes, while the voltage drop across the channel is measured between probes labeled V$_1$ and V$_2$.

We first characterize a typical double-layer WS$_2$ device in air, vacuum and after annealing in vacuum and compare these results (Figure 2). In air, the device shows p-type behavior with channel current $I_{ds}$ = 1 nA at gate voltage $V_g$ = -70 V and bias voltage $V_{ds}$ = 1 V. Pumping down the chamber provides access to both branches of conductance, as shown on Figure 2a. During this experiment, we reached a pressure $p = 2.7 \times 10^{-5}$ mbar before annealing (green curve on Figure 2a) and $p = 5 \times 10^{-7}$ mbar after 40 hours annealing at 115°C (blue curve on Figure 2a). As the pressure is decreased, the conductance on the electron side is enhanced, while the p-type conductance is suppressed. Figure 2b shows the threshold voltage $V_{th}$ as a function of pressure with the $V_{th}$ decreasing as the pressure is reduced. In the same time, we notice an improvement of almost two orders of magnitude in the ON current after annealing. This indicates that, while doping is sensitive to pressure, *in-situ* annealing dramatically improves charge carrier injection from the Au contacts. We further analyze the current-voltage $I_{ds}$ - $V_{ds}$ curves at a fixed gate voltage $V_g$ = 60V before (Figure 2c) and after (Figure 2d) annealing. While the curve before annealing in vacuum is asymmetric and hysteretic, the $I_{ds}$-$V_{ds}$ curve after annealing is symmetric, with an increased ON current and reduced hysteresis. We have also studied devices with different contacting materials (Ti/Au and Ag/Au, in addition to Au). The results are presented on Figures S1 and S2a in Supplementary information and reveal that the devices with Au contacts show highest ON currents and lowest contact resistance.

The dependence of device performance on vacuum conditions and annealing is in agreement with previous reports on bilayer[31] and monolayer[17] $MoS_2$ devices. In these reports, a similar shift of $V_{th}$ was observed and attributed to the presence of adsorbed $H_2O$ and $O_2$ from the atmosphere and their removal *via in-situ* annealing while recent theoretical calculations predict[32] that the absorption of $H_2$, $O_2$, $H_2O$ on the surface of $MoS_2$ can result in p-type doping. Our observations confirm that $WS_2$ devices also show a performance increase following annealing in vacuum. Their performance in atmospheric conditions is however markedly worse than in the case of $MoS_2$.

In the case of double-layer $WS_2$, we were able to observe both branches of conductance, while after annealing the p-type transport disappeared due to the shift of $V_{th}$. Recovery of n-type doping after *in-situ* annealing in the $WS_2$ flakes, grown by vapor phase transport with Br as a transport agent is in agreement with previous observations of intrinsic doping of synthetically grown[33-34] and natural[14] TMD crystals. The presence of halogen atoms as impurities was recently observed even in natural samples,[14] and might be responsible for the excess of electrons in the inspected flakes.

We now turn to single-layer devices. Figure 3 shows the room-temperature characterization of a single-layer $WS_2$ transistor in vacuum and the effect of *in-situ* annealing. This device has a length of 6.8 μm, width of 2.5 μm and distance between the voltage probes of 3.7 μm. As shown on Figure 2, placing the device in vacuum results in a shift of $V_{th}$ towards negative values, revealing the pristine electronic state of the material, in the absence of strong doping by the adsorbates. Further annealing improves the ON-state current and provides further shift of $V_{th}$. We perform characterization of a chosen device with different annealing times up to 145 hours in vacuum. Figure 3a shows the $I_{ds}$-$V_g$ curves for different annealing times while Figure 3b presents the dependence of the sheet conductivity on the back-gate voltage $V_g$ for different annealing times. On Figure 3c we show the extracted two-terminal and four-terminal resistance for the fixed gate voltage for the same device. Interestingly, while the four-terminal resistance reaches saturation after 60-80 hours of annealing, the two-terminal resistance keeps decreasing. We also extracted the field-effect mobility $\mu_{FE}$ from our four-terminal measurements, using the expression $\mu_{FE} = \left[ d\sigma / dV_g \right] \times \left[ 1 / C_{ox} \right]$, where σ is the sheet conductivity of the channel and $C_{ox}$ = 1.28×10$^{-8}$ F·cm$^{-2}$ the geometric back-gate capacitance per unit area. The values of two-terminal and four-terminal mobility are presented in Figure 3d. While the four-terminal mobility already saturates at $\mu_{FE}$ ~ 40 cm$^2$V$^{-1}$s$^{-1}$ after 60 hours of annealing, the two-terminal mobility continues to improve gradually. This is in line with our previous observation that the annealing does not influence doping but improves the contact resistance.

Figure 3e presents the extracted value of four-terminal field-effect mobility as a function of the back-gate voltage $V_g$ after the last annealing step (145 hours). The values are extracted by numerically differentiating the conductivity, thus introducing noise in the data. We can see that the mobility *vs.* $V_g$ reaches a plateau at high $V_g$, which corresponds to a linear dependence of σ on $V_g$. This behavior is similar to recent observations in $MoS_2$ and is an indication of device performance limited by short-range scattering.[35] Dashed lines provide a reference for the minimum and maximum values of mobility in this regime. From this curve, we extract a value for the room-temperature mobility of $\mu_{FE} = 50 \pm 7$ $cm^2V^{-1}s^{-1}$. To evaluate the accuracy of these measurements, we show on Figure 3f the drain current as a function of the voltage drop between the two voltage probes ($V_1$ and $V_2$). The dependence is linear in the entire range of applied bias voltage $V_{ds}$ (±2V) for the range of gate voltages $V_g$ where we extract the field-effect mobility ($V_g$ ~ 60 - 80 V). This confirms that the four-terminal conductivity measurements are reliable and not influenced by Schottky barriers that could be present at the voltage probes. Finally, we estimate the $I_{on}/I_{off}$ ratio for this device after the final step of annealing. The ratio $I_{on}/I_{off}$ ~ $10^6$, estimated with an applied bias of $V_{ds}$ = 1 V (Figure S3b), is larger than previously reported values in the case of multilayer $WS_2$ (ref. 26) because of the enhanced electrostatic control achieved in single-layers.

**Temperature-dependent electrical transport.** After investigating single and double-layer $WS_2$ devices in vacuum and at room temperature, we move on to the temperature-dependent study of electronic transport. The devices were annealed in the cryogenic setup until the saturation of the four-terminal mobility could be reached. The four-terminal sheet conductivities as a function of gate voltage $V_g$ and temperature are presented on Figure 4a in the form of a 2D map.

A crossover from an insulating regime at low gate voltages (decrease of conductivity with decreasing temperature) to a metallic regime (increase of conductivity with decreasing temperature) is observed around $V_g$ ~ 70 V (see region in dashed circle on Figure 4b). Similar observations have been made in the case of $MoS_2$.[15] Figure 4c presents the values of sheet conductivity in units of $e^2/h$, where $e$ is the elementary charge and $h$ the Planck constant. The crossover region between insulating and metallic states corresponds to a value of the sheet conductivity slightly lower than $e^2/h$. This is consistent with values reported in top-gated[15] as well as back-gated[16] $MoS_2$ devices.

We now investigate the temperature dependence of the mobility in the metallic regime, presented on Figure 4d. We plot here the four-terminal mobility *vs.* temperature $T$ for the single-layer and double-layer device. As the temperature is lowered from 220 K, the field-effect mobility for the single-layer device (double-layer) increases until ≈83 K, where it saturates at 120-140 $cm^2V^{-1}s^{-1}$ (above 300 $cm^2V^{-1}s^{-1}$ for the double-layer). In the high-temperature part (83-220 K), the dependence of the mobility on the temperature follows a power-law dependence $\mu_{FE} \propto T^{-\gamma}$, with a temperature damping factor γ = 0.73 (fits are shown as black lines on Figure 4d). This value is similar to that reported in encapsulated single-layer $MoS_2$ (ref. 15) and smaller than in unencapsulated single-layer $MoS_2$.[16] This temperature dependence has been ascribed to a complex interplay between homopolar phonon mode quenching,[36-37] temperature-dependent screening[38] and charged-impurity scattering.[39] For the double-layer device, the power-law fit results in the damping factor γ = 1.75, indicating fundamental differences in interaction with phonons and charged impurities in comparison to single-layer case.

**Insulating regime.** The behaviour in the insulating regime can be studied in the frameworks of thermally activated and variable-range hopping (VRH) models. First, from the high-temperature activation behaviour (inset of Figure 5a) we can extract the activation energy (Figure 5a) by fitting the sheet conductivity with the expression $G(T) = G_0 e^{-E_a/k_B T}$ where $G_0$ is a constant, $E_a$ the activation energy $k_B$ the Boltzmann constant and $T$ the temperature. This activation energy corresponds to the thermal activation of charge carriers at the Fermi energy into the conduction band. Its dependence on gate voltage thus gives us the dependence of the Fermi energy on gate voltage, *i.e.* the density of states (DOS) just below the conduction band edge. The variation of the Fermi energy $E_F$ with the gate voltage $V_g$ is $dE_F/dV_g = C_{ox}/(C_{ox} + C_t)$, where $C_{ox}$ is the back-gate capacitance and $C_t = e^2 D(E)$ the quantum capacitance, from which the DOS $D(E)$ could be extracted.

This DOS is shown in Figure 5b. As the gate voltage is decreased, we see that the DOS first assumes a constant value below $D_{2D}^{exp} \approx 2 \times 10^{14} eV^{-1} cm^{-2}$, close to the expected value for the DOS in the conduction band of WS$_2$: $D = g_v \times m^*/\pi\hbar^2$,[40] where $g_v$=2 stands for the valley degeneracy and $m^*$ is the electron effective mass, which we take from recent theoretical calculations to be $m^* = 0.34\ m_0$.[19] We find $D_{2D}^{theory} = 2.85 \times 10^{14} eV^{-1} cm^{-2}$. At lower gate voltages, the DOS exhibits an exponential decrease (dashed line in Figure 5b). Such an exponential tail near the conduction band edge has been recently observed in MoS$_2$.[41] Notice that the drop in the DOS corresponds to the room-temperature threshold voltage measured from the linear extrapolation of the output curve to zero, as shown in the inset of Figure 5b. The values of DOS ($D \approx 10^{13} eV^{-1} cm^{-2}$) measured around V$_g$ = 0V on the other hand are comparable with the ones obtained with the same model for monolayer MoS$_2$,[41] which is in general attributed to the density of trap states inside the gap.

In the lower temperature regime (temperature between 20 K and 250 K), variable range hopping (VRH) model,[42-43] characterized by the expression $\sigma \propto e^{-(T_0/T)^{1/3}}$ can be used to describe electrical transport in the WS$_2$ device. Using the VRH model, one can extract the localization length $\xi_{loc} = \sqrt{13.8/k_B DT}$ where $D$ is the density of states.[44] Using the values of $D$ extracted from the high-temperature regime (Figure 5b), together with the values of $T_0$ extracted from the VRH model (Figure 5c), we can determine the localization length as a function of the gate voltage. As shown on Figure 5d, this value is constant and close to 2 nm while the $E_F$ stays inside the bandgap, in agreement with a recent report in few-layer WS$_2$.[30] As the gate voltage is increased above the threshold voltage, the localization length starts to linearly increase, indicating the onset of electron delocalization.

**CONCLUSIONS**

In conclusion, by removing adsorbates and atmospheric contaminants using *in-situ* annealing, we could study the electronic properties of pristine single-layer and double-layer WS$_2$. Our measurements show that WS$_2$ is a 2D semiconductor with promising electronic properties. Field-effect mobility 50 ± 7 cm$^2$V$^{-1}$s$^{-1}$ at room temperature and current modulation I$_{on}$/I$_{off}$ ~10$^6$ are comparable to those of single-layer MoS$_2$. The annealing experiments allowed us to outline the important differences between adsorbate removal and contact annealing. The careful annealing procedure also allowed low-temperature transport measurements to be performed in a four-terminal configuration. We observed a crossover between an insulating and a metallic behavior at high charge densities. The mobility shows a power law dependence on temperature and saturates below 83 K at 140 cm$^2$V$^{-1}$s$^{-1}$ for the single-layer case and above 300 cm$^2$V$^{-1}$s$^{-1}$ for the double-layer case. Transport in the insulating regime was modeled as well, which allowed us to observe the same exponential band-tail in single-layer WS$_2$, as the one recently inspected in single-layer MoS$_2$,[41] as well as to extract the localization length. Our results indicate the high quality of presented WS$_2$ field-effect transistors, on par with similar MoS$_2$ and WSe$_2$-based devices.[45,46] The back-gated device configuration used in this work can be further improved by using high-k dielectrics and top gates, which will be the subject of future work.

**METHODS**

WS$_2$ single crystals were grown using the vapor-phase transport method,[47] with Br as the transport agent. Ultrathin WS$_2$ flakes were obtained by micromechanical cleavage of these crystals on degenerately doped n++ silicon covered with 270 nm thermally grown SiO$_2$. The thickness of the crystals was determined by non-contact mode Atomic Force Microscopy (AFM) (Figure 1a), and correlated with optical contrast between the crystal and the substrate. Single and double-layer WS$_2$ crystals were located and six-terminal devices were fabricated using electron-beam lithography (EBL), followed by thermal evaporation of 90 nm thick Au contacts. Other metals, in particular, Ti/Au (2/50 nm) and Ag/Au (5/45nm) stacks were studied as well with results is presented in supplementary information, Figure S1. The samples were subsequently annealed to reduce the contact resistance in Ar/H$_2$ flow.[4] Some devices with irregular shape were further patterned using a second step of EBL, followed by O$_2$ plasma etching. Several steps of thermal annealing were performed in vacuum at a base pressure of 10$^{-6}$ mbar to remove adsorbates and approach the pristine state of the material. Four-terminal electrical measurements were performed to remove the influence of contact resistance.

AFM imaging is performed using an Asylum Research Cypher AFM. After Au contact deposition, devices are annealed in 100 sccm of Ar and 10 sccm H$_2$ flow at 200 °C for 2 h. Electrical characterization is carried out using Agilent E5270B parameter analyzer and a home-built vacuum annealing setup at a base pressure of 10$^{-6}$ mbar. Cryogenic measurements were performed in an Oxford Instruments Heliox cryo-magnetic system.

**ACKNOWLEDGEMENTS**

We thank S. Bertolazzi (EPFL), D. Lembke (EPFL) and T. Heine (Jacobs University) for useful discussions. Device fabrication was carried out in the EPFL Center for Micro/Nanotechnology (CMI). We thank Z. Benes (CMI) for technical support with e-beam lithography. This work was financially supported by the European Union's Seventh Framework Programme FP7/2007-2013 under Grant Agreement No. 318804 (SNM) and Swiss SNF Sinergia Grant no. 147607.

# REFERENCES

1. Novoselov, K. S.; Geim, A. K.; Morozov, S. V.; Jiang, D.; Zhang, Y.; Dubonos, S. V.; Grigorieva, I. V.; Firsov, A. A., Electric Field Effect in Atomically Thin Carbon Films. *Science* 2004, 306, 666-669.
2. Bolotin, K. I.; Sikes, K. J.; Jiang, Z.; Klima, M.; Fudenberg, G.; Hone, J.; Kim, P.; Stormer, H. L., Ultrahigh electron mobility in suspended graphene. *Solid State Commun.* 2008, 146, 351-355.
3. Wang, Q. H.; Kalantar-Zadeh, K.; Kis, A.; Coleman, J. N.; Strano, M. S., Electronics and optoelectronics of two-dimensional transition metal dichalcogenides. *Nat. Nanotechnol.* 2012, 7, 699-712.
4. Radisavljevic, B.; Whitwick, M. B.; Kis, A., Integrated Circuits and Logic Operations Based on Single-Layer $MoS_2$. *ACS Nano* 2011, 5, 9934-9938.
5. Wang, H.; Yu, L.; Lee, Y.-H.; Shi, Y.; Hsu, A.; Chin, M. L.; Li, L.-J.; Dubey, M.; Kong, J.; Palacios, T., Integrated Circuits Based on Bilayer $MoS_2$ Transistors. *Nano Lett.* 2012, 12, 4674-4680.
6. Radisavljevic, B.; Whitwick, M. B.; Kis, A., Small-signal amplifier based on single-layer $MoS_2$. *Appl. Phys. Lett.* 2012, 101, 043103.
7. Bertolazzi, S.; Krasnozhon, D.; Kis, A., Nonvolatile Memory Cells Based on $MoS_2$/Graphene Heterostructures. *ACS Nano* 2013, 7, 3246-3252.
8. Pu, J.; Yomogida, Y.; Liu, K.-K.; Li, L.-J.; Iwasa, Y.; Takenobu, T., Highly Flexible $MoS_2$ Thin-Film Transistors with Ion Gel Dielectrics. *Nano Lett.* 2012, 12, 4013-4017.
9. Lebegue, S.; Eriksson, O., Electronic structure of two-dimensional crystals from ab initio theory. *Phys. Rev. B* 2009, 79, 115409.
10. Splendiani, A.; Sun, L.; Zhang, Y.; Li, T.; Kim, J.; Chim, C.-Y.; Galli, G.; Wang, F., Emerging Photoluminescence in Monolayer $MoS_2$. *Nano Lett.* 2010, 10, 1271-1275.
11. Mak, K. F.; Lee, C.; Hone, J.; Shan, J.; Heinz, T. F., Atomically Thin $MoS_2$: A New Direct-Gap Semiconductor. *Phys. Rev. Lett.* 2010, 105, 136805.
12. Kuc, A.; Zibouche, N.; Heine, T., Influence of quantum confinement on the electronic structure of the transition metal sulfide $TS_2$. *Phys. Rev. B* 2011, 83, 245213.
13. Lopez-Sanchez, O.; Lembke, D.; Kayci, M.; Radenovic, A.; Kis, A., Ultrasensitive photodetectors based on monolayer $MoS_2$. *Nat. Nanotechnol.* 2013, 8, 497-501.
14. Yin, Z.; Li, H.; Li, H.; Jiang, L.; Shi, Y.; Sun, Y.; Lu, G.; Zhang, Q.; Chen, X.; Zhang, H., Single-Layer $MoS_2$ Phototransistors. *ACS Nano* 2012, 6, 74-80.
15. Radisavljevic, B.; Kis, A., Mobility engineering and a metal-insulator transition in monolayer $MoS_2$. *Nat. Mater.* 2013, 12, 815-820.
16. Baugher, B. W. H.; Churchill, H. O. H.; Yang, Y.; Jarillo-Herrero, P., Intrinsic Electronic Transport Properties of High-Quality Monolayer and Bilayer $MoS_2$. *Nano Letters* 2013, 13, 4212-4216.
17. Jariwala, D.; Sangwan, V. K.; Late, D. J.; Johns, J. E.; Dravid, V. P.; Marks, T. J.; Lauhon, L. J.; Hersam, M. C., Band-like transport in high mobility unencapsulated single-layer $MoS_2$ transistors. *Appl. Phys. Lett.* 2013, 102, 173107.
18. Elías, A. L.; Perea-López, N.; Castro-Beltrán, A.; Berkdemir, A.; Lv, R.; Feng, S.; Long, A. D.; Hayashi, T.; Kim, Y. A.; Endo, M., *et al.*, Controlled Synthesis and Transfer of Large-Area $WS_2$ Sheets: From Single Layer to Few Layers. *ACS Nano* 2013, 7, 5235-5242.

19. Liu, L.; Kumar, S. B.; Ouyang, Y.; Guo, J., Performance Limits of Monolayer Transition Metal Dichalcogenide Transistors. *IEEE Trans. El. Dev.* 2011, 58, 3042-3047.
20. Zhu, Z. Y.; Cheng, Y. C.; Schwingenschlögl, U., Giant spin-orbit-induced spin splitting in two-dimensional transition-metal dichalcogenide semiconductors. *Phys. Rev. B* 2011, 84, 153402.
21. Xiao, D.; Liu, G.-B.; Feng, W.; Xu, X.; Yao, W., Coupled Spin and Valley Physics in Monolayers of $MoS_2$ and Other Group-VI Dichalcogenides. *Phys. Rev. Lett.* 2012, 108, 196802.
22. Mak, K. F.; McGill, K. L.; Park, J.; McEuen, P. L., The valley Hall effect in $MoS_2$ transistors. *Science* 2014, 344, 1489-1492.
23. Gutiérrez, H. R.; Perea-López, N.; Elías, A. L.; Berkdemir, A.; Wang, B.; Lv, R.; López-Urías, F.; Crespi, V. H.; Terrones, H.; Terrones, M., Extraordinary Room-Temperature Photoluminescence in Triangular $WS_2$ Monolayers. *Nano Lett.* 2012, 13, 3447–3454.
24. Elías, A. L.; Perea-López, N.; Castro-Beltrán, A.; Berkdemir, A.; Lv, R.; Feng, S.; Long, A. D.; Hayashi, T.; Kim, Y. A.; Endo, M., *et al.*, Controlled Synthesis and Transfer of Large-Area $WS_2$ Sheets: From Single Layer to Few Layers. *ACS Nano* 2013, 7, 5235-5242.
25. Lee, Y.-H.; Yu, L.; Wang, H.; Fang, W.; Ling, X.; Shi, Y.; Lin, C.-T.; Huang, J.-K.; Chang, M.-T.; Chang, C.-S., *et al.*, Synthesis and Transfer of Single-Layer Transition Metal Disulfides on Diverse Surfaces. *Nano Lett.* 2013, 13, 1852-1857.
26. Hwang, W. S.; Remskar, M.; Yan, R.; Protasenko, V.; Tahy, K.; Chae, S. D.; Zhao, P.; Konar, A.; Xing, H.; Seabaugh, A., *et al.*, Transistors with chemically synthesized layered semiconductor $WS_2$ exhibiting $10^5$ room temperature modulation and ambipolar behavior. *Appl. Phys. Lett.* 2012, 101, 013107.
27. Georgiou, T.; Jalil, R.; Belle, B. D.; Britnell, L.; Gorbachev, R. V.; Morozov, S. V.; Kim, Y.-J.; Gholinia, A.; Haigh, S. J.; Makarovsky, O., *et al.*, Vertical field-effect transistor based on graphene-$WS_2$ heterostructures for flexible and transparent electronics. *Nat. Nanotechnol.* 2012, 8, 100-103.
28. Braga, D.; Gutiérrez Lezama, I.; Berger, H.; Morpurgo, A. F., Quantitative Determination of the Band Gap of $WS_2$ with Ambipolar Ionic Liquid-Gated Transistors. *Nano Lett.* 2012, 12, 5218-5223.
29. Jo, S.; Ubrig, N.; Berger, H.; Kuzmenko, A. B.; Morpurgo, A. F., Mono- and Bilayer $WS_2$ Light-Emitting Transistors. *Nano Lett.* 2014, 14, 2019–2025.
30. Withers, F.; Bointon, T. H.; Hudson, D. C.; Craciun, M. F.; Russo, S., Electron transport of $WS_2$ transistors in a hexagonal boron nitride dielectric environment. *Sci. Rep.* 2014, 4, 4967.
31. Qiu, H.; Pan, L.; Yao, Z.; Li, J.; Shi, Y.; Wang, X., Electrical characterization of back-gated bi-layer $MoS_2$ field-effect transistors and the effect of ambient on their performances. *Appl. Phys. Lett.* 2012, 100, 123104.
32. Yue, Q.; Shao, Z.; Chang, S.; Li, J., Adsorption of gas molecules on monolayer $MoS_2$ and effect of applied electric field. *Nanoscale Research Letters* 2013, 8, 425.
33. El-Mahalawy, S. H.; Evans, B. L., Temperature dependence of the electrical conductivity and hall coefficient in 2H-$MoS_2$, $MoSe_2$, $WSe_2$, and $MoTe_2$. *Phys. Status Solidi B* 1977, 79, 713-722.
34. Fivaz, R.; Mooser, E., Mobility of Charge Carriers in Semiconducting Layer Structures. *Physical Review* 1967, 163, 743-755.

35. Schmidt, H.; Wang, S.; Chu, L.; Toh, M.; Kumar, R.; Zhao, W.; Castro Neto, A. H.; Martin, J.; Adam, S.; Özyilmaz, B.*, et al.*, Transport Properties of Monolayer $MoS_2$ Grown by Chemical Vapor Deposition. *Nano Lett.* 2014.
36. Kaasbjerg, K.; Thygesen, K. S.; Jacobsen, K. W., Phonon-limited mobility in n-type single-layer $MoS_2$ from first principles. *Phys. Rev. B* 2012, 85, 115317.
37. Kaasbjerg, K.; Thygesen, K. S.; Jauho, A.-P., Acoustic phonon limited mobility in two-dimensional semiconductors: Deformation potential and piezoelectric scattering in monolayer $MoS_2$ from first principles. *Phys. Rev. B* 2013, 87, 235312.
38. Ma, N.; Jena, D., Charge Scattering and Mobility in Atomically Thin Semiconductors. *Phys. Rev. X* 2014, 4, 011043.
39. Ong, Z.-Y.; Fischetti, M. V., Mobility enhancement and temperature dependence in top-gated single-layer $MoS_2$. *Phys. Rev. B* 2013, 88, 165316.
40. Kim, S.; Konar, A.; Hwang, W.-S.; Lee, J. H.; Lee, J.; Yang, J.; Jung, C.; Kim, H.; Yoo, J.-B.; Choi, J.-Y.*, et al.*, High-mobility and low-power thin-film transistors based on multilayer $MoS_2$ crystals. *Nat. Commun.* 2012, 3, 1011-1017.
41. Zhu, W.; Low, T.; Lee, Y.-H.; Wang, H.; Farmer, D. B.; Kong, J.; Xia, F.; Avouris, P., Electronic transport and device prospects of monolayer molybdenum disulphide grown by chemical vapour deposition. *Nat. Commun.* 2014, 5.
42. Lösche, A.; Mott, N. F.; Davis, E. A., Electronic Processes in Non-Crystalline Materials. *Kristall und Technik* 1972, 7, K55-K56.
43. Van Keuls, F. W.; Hu, X. L.; Jiang, H. W.; Dahm, A. J., Screening of the Coulomb interaction in two-dimensional variable-range hopping. *Phys. Rev. B* 1997, 56, 1161-1169.
44. Ghatak, S.; Pal, A. N.; Ghosh, A., Nature of Electronic States in Atomically Thin $MoS_2$ Field-Effect Transistors. *ACS Nano* 2011, 5, 7707-7712.
45. Fang, H.; Chuang, S.; Chang, T. C.; Takei, K.; Takahashi, T.; Javey, A., High-Performance Single Layered $WSe_2$ p-FETs with Chemically Doped Contacts. *Nano Lett.* 2012, 12, 3788-3792.
46. Liu, W.; Kang, J.; Sarkar, D.; Khatami, Y.; Jena, D.; Banerjee, K., Role of Metal Contacts in Designing High-Performance Monolayer n-Type $WSe_2$ Field Effect Transistors. *Nano Lett.* 2013, 13, 1983-1990.
47. Agarwal, M. K.; Nagi Reddy, K.; Patel, H. B., Growth of tungstenite single crystals by direct vapour transport method. *J. Cryst. Growth* 1979, 46, 139-142.

**FIGURES**

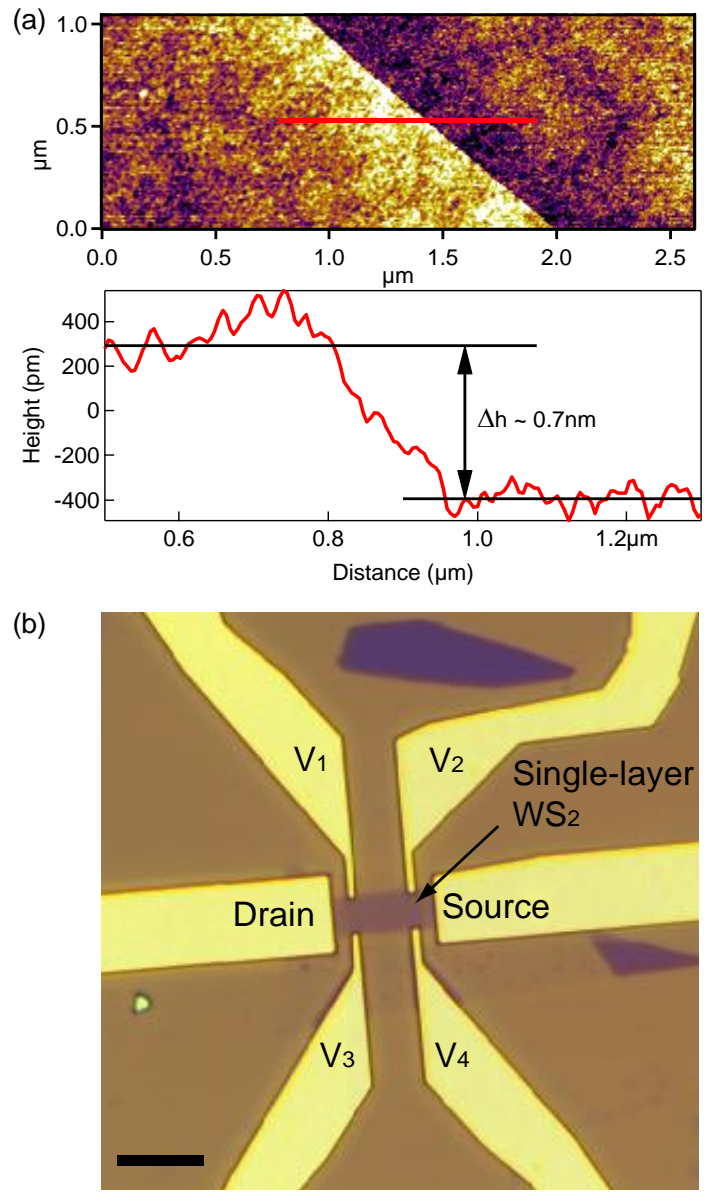

**Figure 1. Fabrication of single-layer WS$_2$ transistors.** (a) AFM image of single-layer WS$_2$ exfoliated on 270 nm of SiO$_2$ and the corresponding height profile (along the red line), from which a height of 0.7 nm can be extracted. Black lines are guides for the eyes indicating the step height (b) Optical micrograph of the device presented in this manuscript based on single-layer WS$_2$ in the Hall bar geometry. Device length $L$ = 6.8 µm and width $W$ = 2.5 µm. Scale bar is 5 µm long.

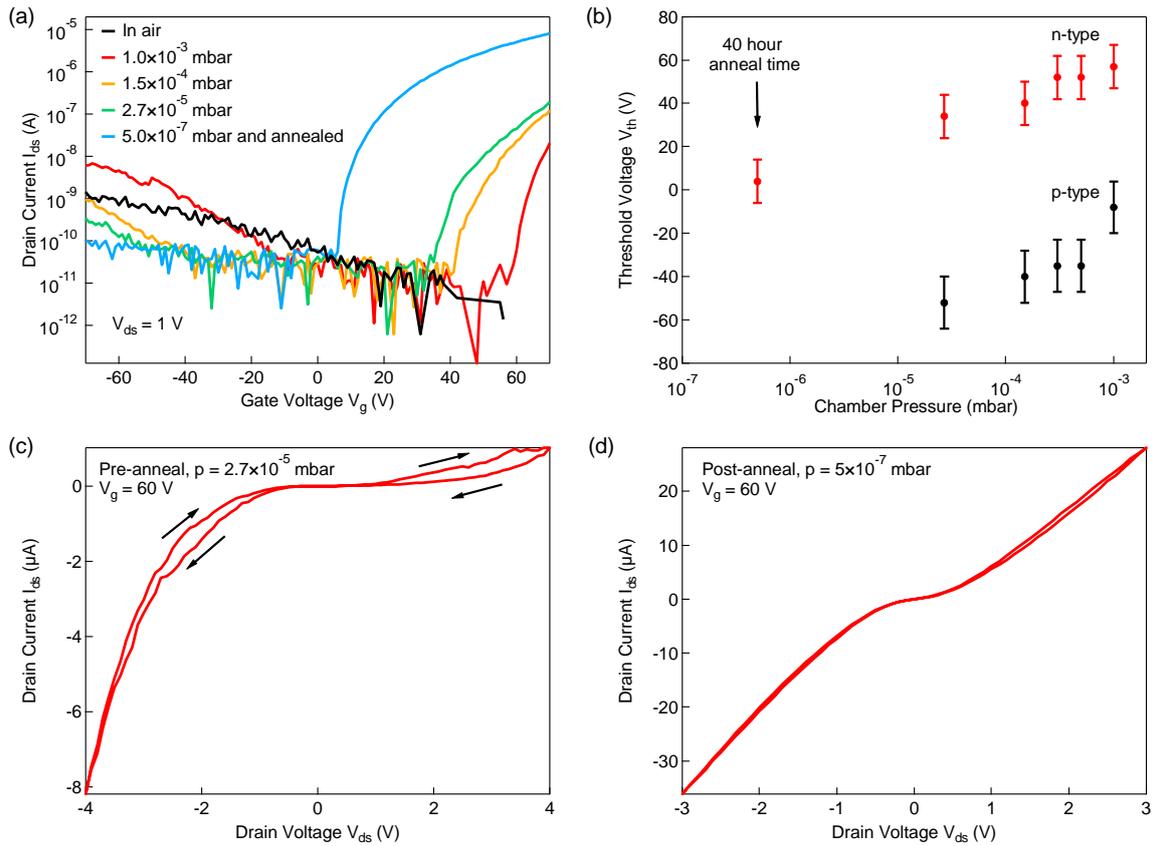

**Figure 2. Characterization of double-layer WS$_2$ transistor in vacuum and the effect of annealing.** (a) Current as a function of gate voltage ($I_{ds}$-$V_g$) in air, vacuum and after annealing. (b) Threshold voltage $V_{th}$ for p-type and n-type conduction as a function of pressure in the system. (c) The dependence of device current $I_{ds}$ on the bias voltage $V_{ds}$ ($I_{ds}$-$V_{ds}$) for a fixed value of gate voltage $V_g = 60$ V before annealing in the vacuum chamber. (d) Same as (c), after annealing for 40 hours at 115 °C.

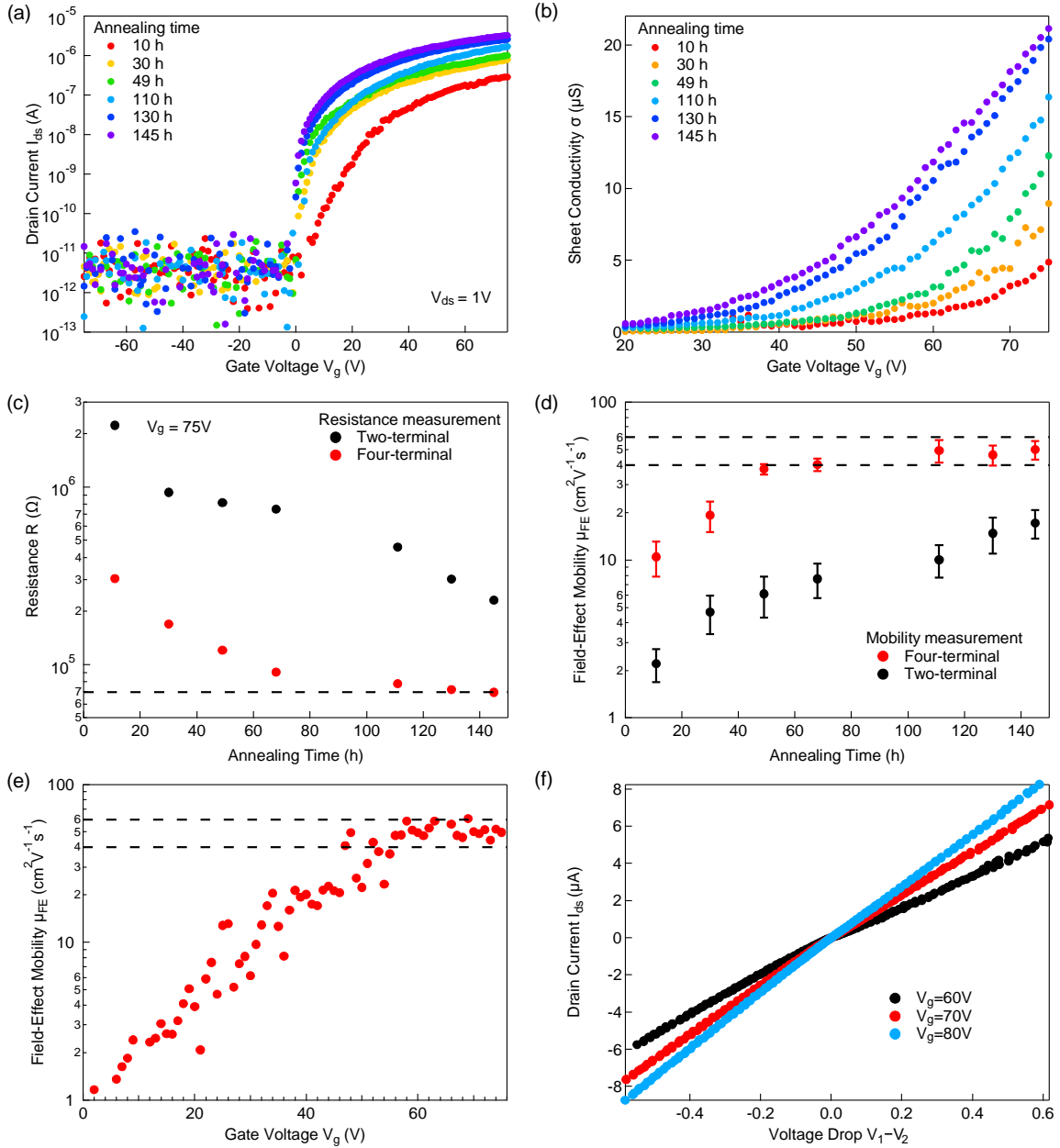

**Figure 3. Performance of the single-layer WS$_2$ transistor in vacuum.** (a) Current-voltage $I_{ds}$-$V_g$ curves for the device recorded at different stages during the annealing process. (b) Sheet conductivity σ as a function of gate voltage $V_g$ for the same annealing times as in (a). (c) Resistance as a function of annealing time at fixed gate voltage ($V_g$=75V). Red markers - four-terminal measurements, black markers - two-terminal measurements. Four-terminal resistance reaches a plateau at ~70 kΩ indicated by the dashed line after 60 hours of annealing, while the two-terminal resistance continues to decrease during the entire annealing procedure. (d) Field-effect mobility of single-layer WS$_2$ *vs.* the annealing time. Red markers - four-terminal mobility, black markers - two-terminal mobility. Dashed lines are given as a guide for the range of extracted mobilities. (e) Room- temperature four-terminal mobility, extracted from the red curve in (a) after 145 hours of annealing. Dashed lines indicate the range of mobility achieved in the linear regime (σ ~ $V_g$). (f) Current *vs.* voltage drop between the voltage probes after 145 h of annealing in vacuum. The linear dependence implies the correct measurements of four-terminal mobility

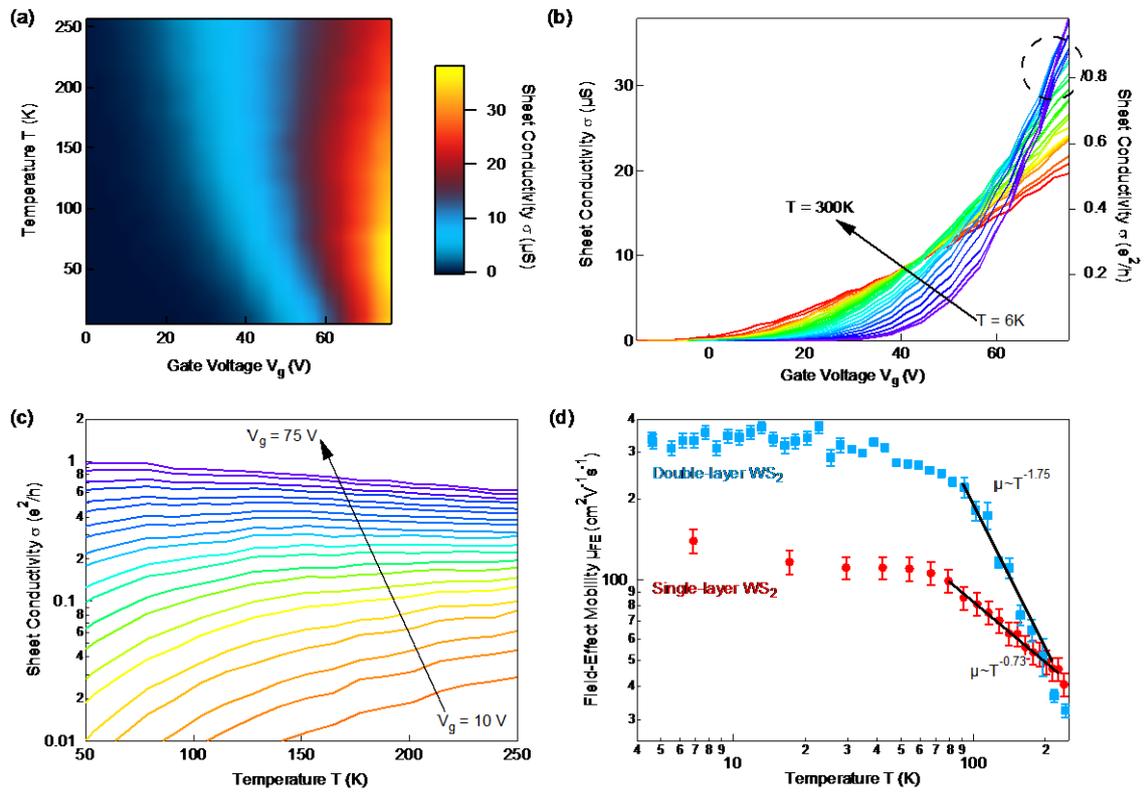

**Figure 4. Temperature-dependent electrical characterization of the single-layer $WS_2$ device.** (a) Dependence of the four-terminal sheet conductivity on gate voltage and temperature. The lower left corner exhibits the maximum sheet conductivity, corresponding to high gate voltages and low temperatures. (b) Sheet conductivity *vs.* gate voltage ($V_g$) for different temperatures. The area highlighted with the dashed circle represents the crossover region from the insulating ($V_g < 70$ V) to the metallic regime ($V_g > 70$V) at low temperatures (between 5 K and 80 K). (c) Dependence of the sheet conductivity in units of $e^2/h$ on the temperature for different gate voltages. (d) Four-terminal field-effect mobility *vs* temperature extracted from the gate voltage range between 65 V and 75 V. Red markers correspond to single-layer, blue - double-layer $WS_2$. The solid black lines are fits to the model $\mu \sim T^{-\gamma}$ in the 83-220 K temperature range.

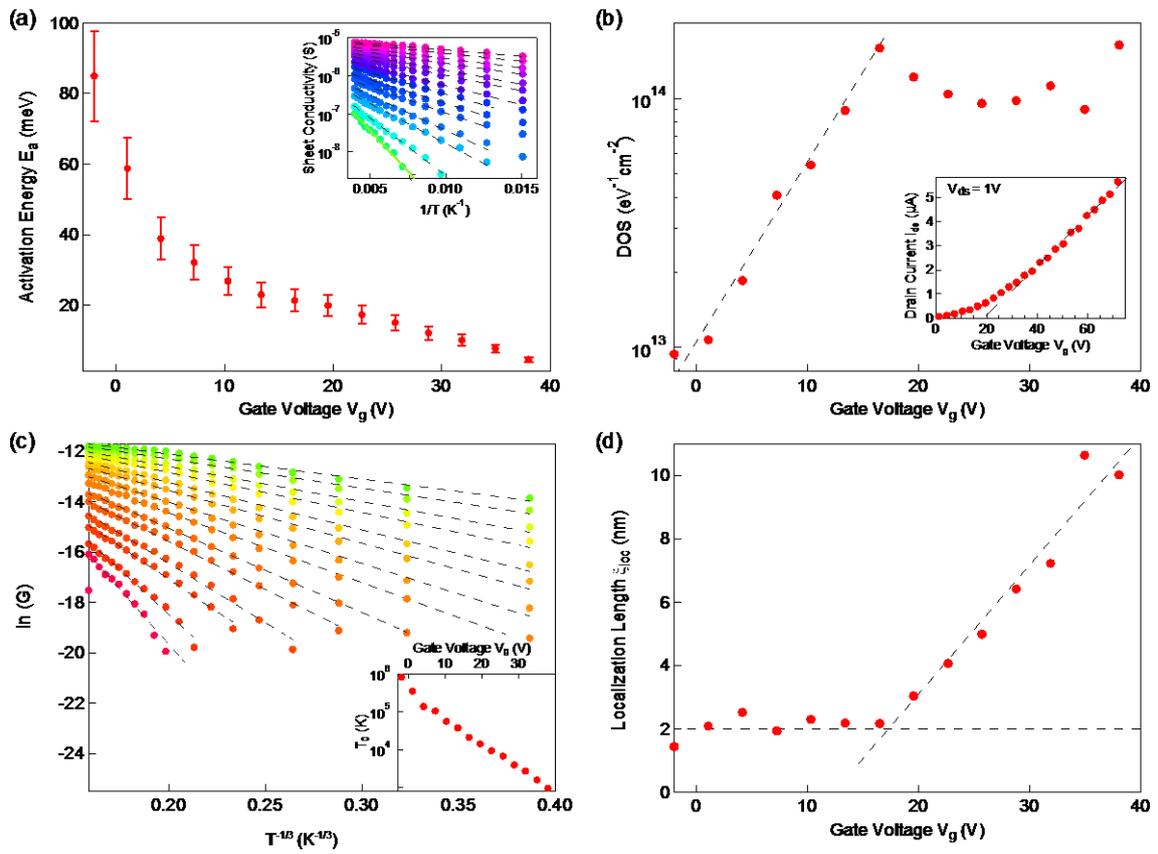

**Figure 5. Electrical transport in the insulating regime of single-layer WS$_2$.** (a) Activation energy $E_a$ as a function of gate voltage $V_g$ extracted from the Arrhenius plots of the sheet conductivity (inset) in the high temperature range. (b) Density of states extracted from the activation energy. Inset: $I_{ds}$-$V_g$ curve for the same device recorded at room temperature with linear extraction of $V_{th}$ (dashed line). (c) Logarithm of sheet conductivity plotted as a function of $T^{-1/3}$ for different gate voltages. The variable range hopping parameter $T_0$ (shown in inset) can be extracted from the slopes of the line fits. (d) The dependence of localization length on the gate voltage $V_g$. Dashed lines are guides for the eye.

# Supplementary Information - Electrical Transport Properties of Single-Layer $WS_2$

Dmitry Ovchinnikov[1], Adrien Allain[1], Ying-Sheng Huang[2], Dumitru Dumcenco[1], Andras Kis[1*]

[1]Electrical Engineering Institute, Ecole Polytechnique Federale de Lausanne (EPFL), CH-1015 Lausanne, Switzerland
[2]Department of Electronic Engineering, National Taiwan University of Science and Technology, Taipei 106, Taiwan

*Correspondence should be addressed to: Andras Kis, andras.kis@epfl.ch

## PERFORMANCE OF DEVICES IN VACUUM

First, we compare the results of devices with different types of metal contact after annealing in vacuum. On Figure S1, the characteristics of several devices of each type is presented, starting from the threshold voltage $V_{th}$ (Figure S1a). All devices were annealed for up to 30 hours. Further annealing did not provide significant change in the threshold voltage for Ag/Au or Ti/Au devices. We also compare the contact resistances of devices normalized for the length of contact area. We fix the difference between gate voltage $V_g$ at which the contact resistance is extracted and the $V_{th}$ to be 70V for each device. These measurements are presented for various annealing times up to 40 hours. Au contacts provided the best results here and allowed correct voltage reading and accurate extraction of four-terminal mobility, as as shown in the main text in Figure 3f. In contrast to Au contacts, Ti/Au and Ag/Au contacts showed higher contact resistance and asymmetric $I_{ds}$-$V_{ds}$ curves. The examples are provided for 72 hours of annealing time on Figure S1c and S1d, where we fixed the back gate voltage and swept the drain-source voltage. An example of a symmetric curve obtained for Au contacts is shown in the main text on Figure 2d.

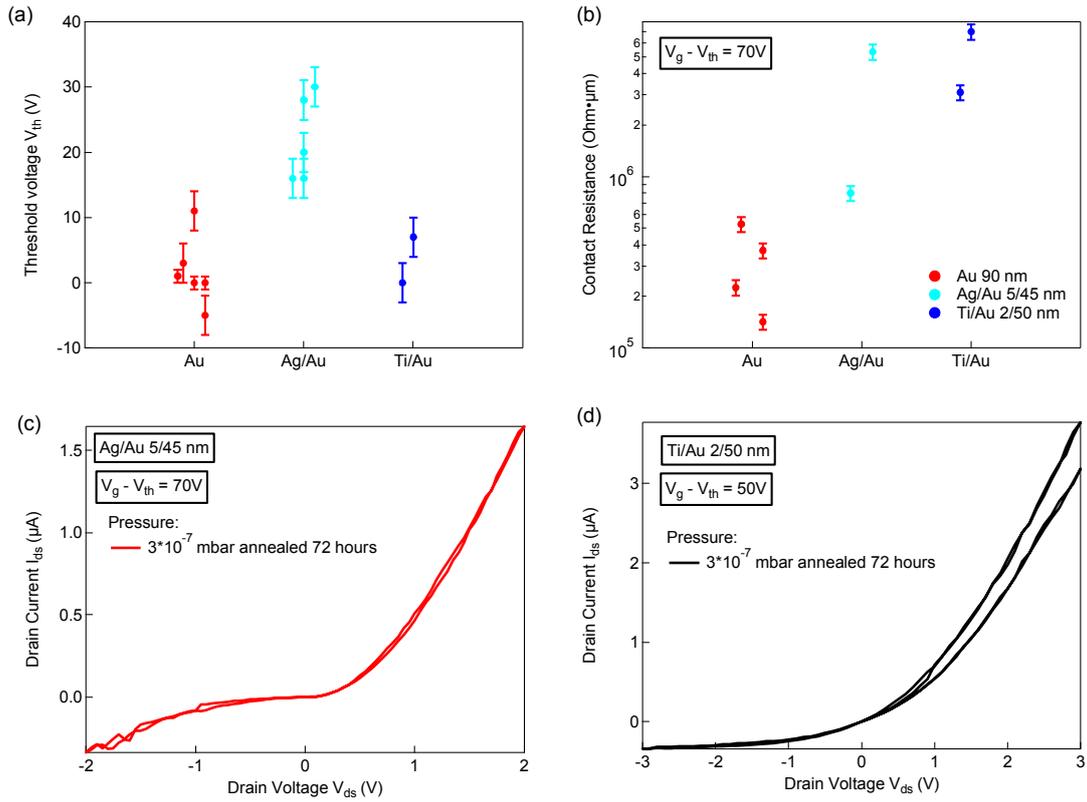

**Figure S1. Comparison of different contact metals.** (a) Dependence of the threshold voltage $V_{th}$ on the metal used for contacts contact for different devices. Measurements were performed after 20-30 hours of annealing in vacuum. (b) Extracted contact resistance for the selected devices at fixed overdrive voltage $V_g - V_{th}$ =70V. (c) Current-voltage ($I_{ds}$-$V_{ds}$) curves at fixed gate voltage for Ag/Au device annealed for 72 hours. (d) Current-voltage ($I_{ds}$-$V_{ds}$) curves at fixed gate voltage for Ti/Au device annealed for 72 hours, double sweep is presented.

On Figure S2, we present the results for mobility of 8 different devices and its dependence on annealing time, as well as extraction of the $I_{on}/I_{off}$ ratio for device ML1.

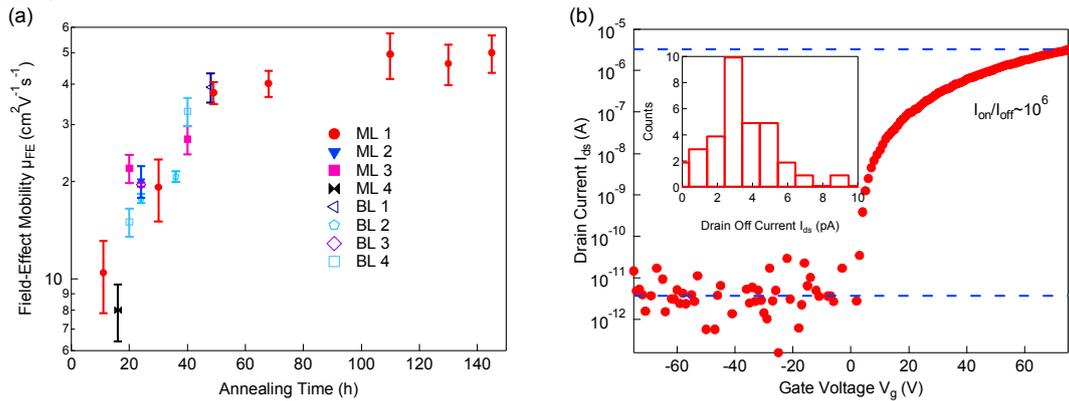

**Figure S2. WS$_2$ devices in vacuum.** (a) Four-terminal field-effect mobility as a function of annealing time for 8 WS$_2$ devices. Filled markers - single-layer, empty markers - double-layer devices. (b) Current-voltage ($I_{ds}$-$V_g$) characteristics of device presented in Figure 3 of main text after 145 hours of annealing in vacuum, $V_{ds}$ = 1V, $I_{on}$ ~ 3µA, $I_{off}$ ~ 3 pA. Inset: histogram showing the distribution of recorded off-state currents, used to estimate the current on/off ratio.